\documentclass[a4paper,11pt]{article}
\usepackage{pos}

\title{$|V_{us}|$ from kaon semileptonic form factor in $N_f = 2+1$ QCD at the physical point on (10 fm)$^4$}
\ShortTitle{$|V_{us}|$ from kaon semileptonic form factor in $N_f = 2+1$ QCD}

\author*[a,b]{Takeshi Yamazaki}

\author[c]{Ken-ichi~Ishikawa}

\author[b]{Naruhito~Ishizuka}

\author[b]{Yoshinobu~Kuramashi}

\author[d]{Yusuke~Namekawa}

\author[b]{Yusuke~Taniguchi}

\author[a]{Naoya~Ukita}

\author[b]{Tomoteru~Yoshi\'e}

\affiliation[]{\normalsize{\bf \sffamily \hspace{50mm} (PACS Collaboration)}}

\affiliation[a]{Institute of Pure and Applied Sciences, University of Tsukuba, Tsukuba, Ibaraki 305-8571, Japan}

\affiliation[b]{Center for Computational Sciences, University of Tsukuba, Tsukuba, Ibaraki 305-8577, Japan}

\affiliation[c]{Core of Research for the Energetic Universe, Graduate School of Advanced Science and Engineering, Hiroshima University, Higashi-Hiroshima, 739-8526, Japan}

\affiliation[d]{Education and Research Center for Artificial Intelligence and Data Innovation, Hiroshima University, Higashi-Hiroshima 739-8521, Japan}

\emailAdd{yamazaki@het.ph.tsukuba.ac.jp}

\abstract{
We present a preliminary result of the kaon semileptonic form factor 
calculated at the smallest lattice spacing in the PACS10 configurations,
whose physical volumes are more than (10 fm)$^4$ at the physical point. 
The configurations were generated using the Iwasaki gauge action and 
$N_f=2+1$ stout-smeared nonperturbatively $O(a)$ improved Wilson quark action 
at the three lattice spacings, 0.085, 0.063, and 0.041 fm. 
The value of $|V_{us}|$ in the continuum limit is estimated from our results 
including the preliminary one. We compare our result of $|V_{us}|$ 
with the previous results and those through the kaon leptonic decay.
}

\FullConference{The 40th International Symposium on Lattice Field Theory (Lattice 2023)\\
July 31st - August 4th, 2023\\
Fermi National Accelerator Laboratory\\}

\begin{document}
\maketitle

\section{Introduction}

The form factor of the kaon semileptonic ($K_{\ell 3}$) decay plays an important
role in determining $|V_{us}|$, which is one of the Cabibbo-Kobayashi-Maskawa
(CKM) matrix elements, through the $K_{\ell 3}$ decay process.
The current values of the CKM matrix elements in 
the first row including $|V_{us}|$
give 2.2 $\sigma$ violation of the CKM unitarity~\cite{ParticleDataGroup:2022pth}.
It could suggest the existence of physics beyond the standard model.
To clarify the violation, more accurate values of the matrix elements
are necessary.

So far in the various lattice QCD calculations~\cite{Dawson:2006qc,Boyle:2007qe,Lubicz:2009ht,Bazavov:2012cd,Boyle:2013gsa,Bazavov:2013maa,Boyle:2015hfa,Carrasco:2016kpy,Aoki:2017spo,Bazavov:2018kjg}, precise values of 
the $K_{\ell 3}$ form factor were reported.
We also calculated the form factor~\cite{PACS:2019hxd,Ishikawa:2022otj} using 
the $N_f = 2+1$ 
PACS10 configurations~\cite{Ishikawa:2018jee,Shintani:2019wai}, 
where the quark masses are tuned to be the physical
ones on large spacetime volumes of more than (10 fm)$^4$.
In this report, we present updates of those calculations, which are
preliminary results of the form factor with the third PACS10
configuration at a finer lattice spacing,
and also the estimated value of $|V_{us}|$ using the result.

\section{Results}

\begin{table}[!b]
\caption{
Simulation parameters of the PACS10 configurations 
at the three lattice spacings.
The bare coupling ($\beta$), lattice size ($L^3\cdot T$),
physical spatial extent ($L$[fm]), lattice spacing ($a$[fm]),
the number of the configurations ($N_{\rm conf}$), and
pion and kaon masses ($m_\pi$, $m_K$) are tabulated.
  \label{tab:sim_param}
}
\begin{center}
\begin{tabular}{cccccccc}\hline\hline
$\beta$ & $L^3\cdot T$ & $L$[fm] & $a$[fm] &
$N_{\rm conf}$ & $m_\pi$[MeV] & $m_K$[MeV]\\\hline
2.20 & 256$^4$ & 10.5 & 0.041 & 20 & 142 & 514 \\
2.00 & 160$^4$ & 10.1 & 0.063 & 20 & 138 & 505 \\
1.82 & 128$^4$ & 10.9 & 0.085 & 20 & 135 & 497 \\\hline\hline
\end{tabular}
\end{center}
\end{table}

\subsection{$K_{\ell 3}$ form factors}

The simulation parameters for the three ensembles of the PACS10 configuration
are tabulated in Table~\ref{tab:sim_param}.
All the configurations were generated using 
the Iwasaki gauge action~\cite{Iwasaki:2011jk}
and a non-perturbative $O(a)$-improved Wilson quark action 
with the six-stout-smeared link~\cite{Morningstar:2003gk}.

The same quark action is employed in the measurements 
for the two- and three-point functions for the $K_{\ell 3}$ form factors.
We use the random source operator spread in the spin, color, and
spatial spaces proposed in Ref.~\cite{Boyle:2008yd}.
The periodic boundary condition is imposed in the spatial directions
in the calculation of the correlation functions,
while in the temporal direction the periodic and anti-periodic
boundary conditions are employed.
The average of the two-point correlation functions with the different boundary
conditions makes the periodicity of the temporal direction effectively doubled.
Furthermore, a similar average of the three-point functions suppresses
the wrapping around effect~\cite{Kakazu:2017fhv}
in the small momentum region~\cite{PACS:2019hxd,Ishikawa:2022otj}.

The three-point functions are calculated using the local weak vector
current and also using the conserved vector one to investigate
the lattice spacing dependence of the form factors.
We utilize the renormalization factor of
the local vector current given by
$Z_V = 1/\sqrt{F^{\rm bare}_\pi(0)F^{\rm bare}_K(0)}$,
where $F^{\rm bare}_H(0)$ for $H = \pi, K$ 
is the bare electromagnetic form factor
evaluated using the local vector current
at the zero momentum transfer squared, $q^2 = 0$.

From the three-point function in each $q^2$, 
the $K_{\ell 3}$ decay matrix element
is extracted, which is expressed by the two form factors,
$f_+(q^2)$ and $f_-(q^2)$, given by
\begin{eqnarray}
\langle \pi (\vec{p}_{\pi}) \left | V_{\mu} \right | K(\vec{p}_{K}) \rangle = ({p}_{K}+{p}_{\pi})_{\mu}f_{+}(q^2)+ ({p}_{K}-{p}_{\pi})_{\mu}f_{-}(q^2),
\label{eq:def_matrix_element}
\end{eqnarray}
where $V_\mu$ is the weak vector current.
Using the two form factors, another form factor $f_0(q^2)$ is defined as,
\begin{eqnarray}
f_{0}(q^2) = f_{+}(q^2) + \frac{-q^2}{{m^2_{K}}-{m^2_{\pi}}}f_{-}(q^2).
\label{eq:f0}
\end{eqnarray}

\begin{figure}[!t]
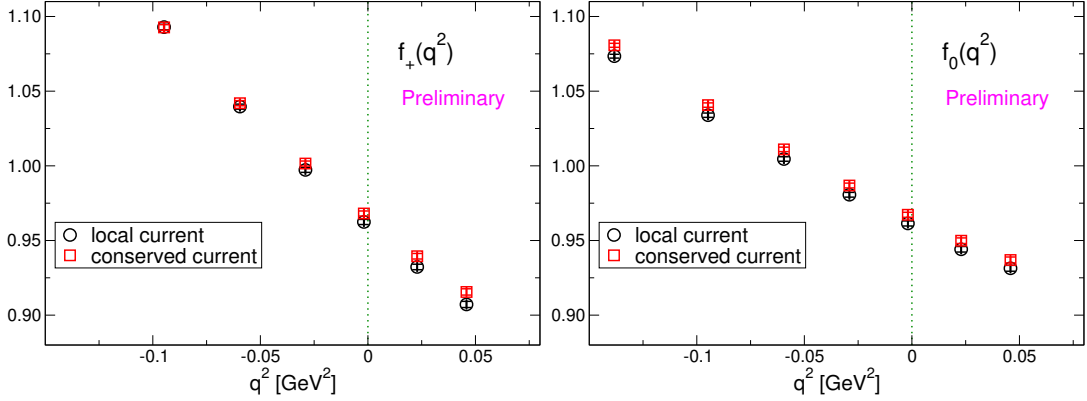

 \centering
 \includegraphics*[scale=0.41]{figs/fplus_b220-con.eps}
 \includegraphics*[scale=0.41]{figs/fzero_b220-con.eps}
 \caption{
Preliminary results for the $K_{\ell 3}$ form factors,
$f_+(q^2)$ (left) and $f_0(q^2)$ (right), as a function of $q^2$
at $a = 0.041$ fm.
The circle and square symbols denote the data using the local and
conserved vector currents, respectively.
  \label{fig:fpz_b220}
 }
\end{figure}

The preliminary results for the $K_{\ell 3}$ form factors,
$f_+(q^2)$ and $f_0(q^2)$, at $a = 0.041$ fm 
are presented in Fig.~\ref{fig:fpz_b220}.
The figure shows that the data from the two vector currents
have clear signals in both the form factors, and the data near $q^2 = 0$ 
can be computed thanks to the large volume of the PACS10 configuration
even with the periodic boundary condition in the spatial directions.
The conserved current data are systematically larger than the
local ones in almost all the $q^2$ regions.
We, however, observe that the discrepancy
between the conserved and local data becomes smaller than
those at the coarser lattice spacings~\cite{Ishikawa:2022otj}.

\begin{figure}[!t]
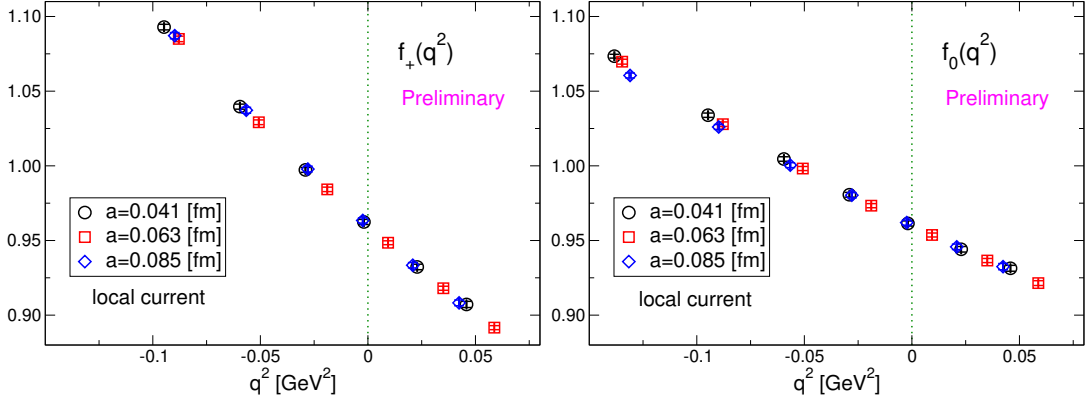

 \centering
 \includegraphics*[scale=0.41]{figs/fplus-lcl_q2.eps}
 \includegraphics*[scale=0.41]{figs/fzero-lcl_q2.eps}
 \caption{
Lattice spacing dependences for $f_+(q^2)$ (left) and $f_0(q^2)$ (right)
as a function of $q^2$ with the local vector current.
The data at $a = 0.041$ fm are preliminary, while other two data
are presented in Ref.~\cite{Ishikawa:2022otj}.
  \label{fig:fpz_adep-loc}
 }
\end{figure}

The preliminary results for the two form factors with the local current 
are compared with the data at the different lattice spacings 
in Fig.~\ref{fig:fpz_adep-loc}.
Although there are tiny discrepancies among the data at the three different 
lattice spacings in the smaller and larger $q^2$ regions,
the data near $q^2 = 0$ can be expressed by
a monotonic function of $q^2$.
It suggests that the local current data seems to have 
a smaller lattice spacing effect.
On the other hand, it is observed that 
the conserved current data have larger lattice
spacing dependence than those in the local data.
%as shown in Fig.~\ref{fig:fpz_adep-con}.
%The data in the $q^2=0$ region have a trend to decrease with 
%the lattice spacing in both the form factors.
%From this trend and the observation that the conserved data
%is systamtcally higher than the local data discussed above,
%it is expected that the conserved data approaches to the local one
%toward the continuum limit.

%\begin{figure}[!t]
% \centering
% \includegraphics*[scale=0.41]{figs/fplus-cns_q2.eps}
% \includegraphics*[scale=0.41]{figs/fzero-cns_q2.eps}
% \caption{
%The same figure as Fig.~\ref{fig:fpz_adep-loc} but for
%the conserved current data.
%  \label{fig:fpz_adep-con}
% }
%\end{figure}

\subsection{$f_+(0)$ in the continuum limit and $|V_{us}|$}

\begin{figure}[!t]
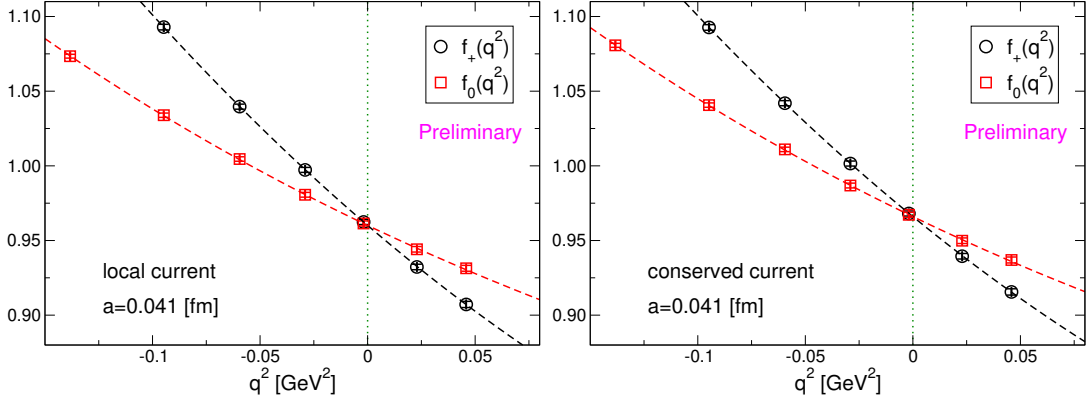

 \centering
 \includegraphics*[scale=0.41]{figs/fplus_fzero_b220-lcl.eps}
 \includegraphics*[scale=0.41]{figs/fplus_fzero_b220-cns.eps}
 \caption{
Fit results for the form factors
using the local (left) and conserved (right) vector currents 
at $a = 0.041$ fm expressed by dashed curves.
The circle and diamond symbols represent $f_+(q^2)$ and $f_0(q^2)$,
respectively.
  \label{fig:q2int}
 }
\end{figure}

A $q^2$ interpolation is carried out to obtain the value of $f_+(0)$,
which is essential to determine the value of $|V_{us}|$ through 
the $K_{\ell 3}$ decay.
In the interpolation we employ fit functions for $f_+(q^2)$ and $f_0(q^2)$
based on the next-to-leading order formulas in
the SU(3) chiral perturbation theory~\cite{Gasser:1984ux,Gasser:1984gg}
with some additional terms and a constraint of $f_+(0) = f_0(0)$.
The fit formulas are given in Ref.~\cite{PACS:2019hxd}.
Figure~\ref{fig:q2int} presents that the fit in each current data
at $a = 0.041$ fm works well.
The values of $f_+(0)$ are obtained from the fits of each current data.
As shown in Table~\ref{tab:sim_param}, the values for $m_\pi$ and $m_k$
are slightly different from the physical ones.
A short chiral extrapolation of $f_+(0)$ to the physical meson masses
is carried out using the same formulas as the fit forms.

\begin{figure}[!t]
 \centering
 \includegraphics*[scale=0.41]{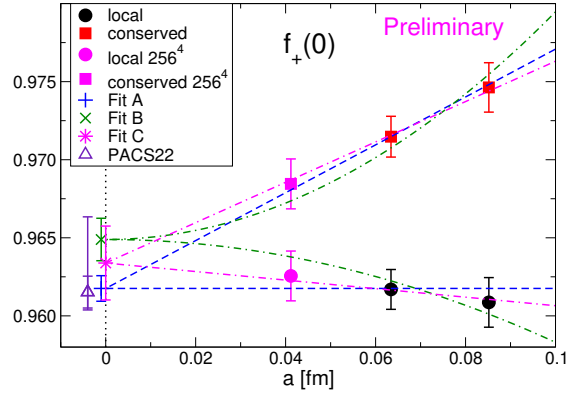}
 \caption{
Lattice spacing dependence of $f_+(0)$.
The circle and square symbols represent the data
with the local and conserved vector currents, respectively.
Different curves correspond to fit results using different fit forms
for the continuum extrapolation.
The triangle symbol denotes our previous result in the continuum limit~\cite{Ishikawa:2022otj}.
  \label{fig:fp0_a0}
 }
\end{figure}

The lattice spacing dependence of $f_+(0)$ is plotted in Fig.~\ref{fig:fp0_a0}.
The magenta symbols are preliminary results at $a = 0.041$ fm obtained
from the above fit.
The other data were determined with the same $q^2$ interpolations 
at each lattice spacing in Ref.~\cite{PACS:2019hxd,Ishikawa:2022otj}.
As discussed in the last subsection,
%observed in Figs.~\ref{fig:fpz_adep-loc} and \ref{fig:fpz_adep-con},
the local current data has an almost flat behavior against the lattice spacing,
while the conserved current data has a clear slope.
In order to estimate $f_+(0)$ in the continuum limit,
we investigate three fit forms using the two current data.
In the fit we assume the results from the two currents coincide
in the continuum limit.
The one fit form utilizes constant and linear functions of $a$ for
the local and conserved vector current data, respectively,
whose result is denoted by Fit A in Fig.~\ref{fig:fp0_a0}.
Fit B and C in the figure adopt quadratic and linear functions of $a$,
respectively, for both the current data.
The two results obtained from Fit A and B differ beyond their statistical
errors, while the result from Fit C covers the other two fit results
within its error.
Thus, we take the result from Fit C as our preliminary result in
this calculation.
The result is consistent with our previous result estimated from
the two coarser lattice spacing data~\cite{Ishikawa:2022otj}.

\begin{figure}[!t]
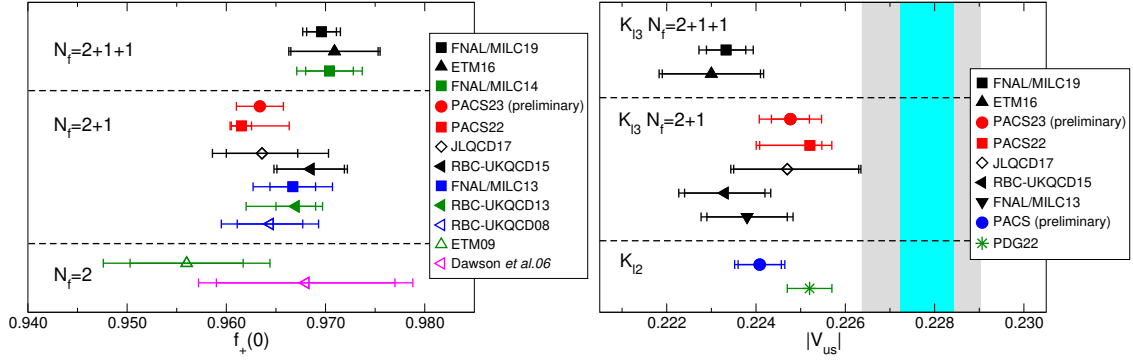

 \centering
 \includegraphics*[scale=0.37]{figs/f00.eps}
 \includegraphics*[scale=0.37]{figs/vus.eps}
 \caption{
{\bf Left :}
Comparison of our preliminary result of $f_+(0)$ with
the previous results~\cite{Dawson:2006qc,Boyle:2007qe,Lubicz:2009ht,Bazavov:2012cd,Boyle:2013gsa,Bazavov:2013maa,Boyle:2015hfa,Carrasco:2016kpy,Aoki:2017spo,Bazavov:2018kjg,Ishikawa:2022otj}.
The inner and outer errors
express the statistical and total errors.
The total error is evaluated by adding the statistical and systematic errors in quadrature.
{\bf Right :}
Comparison of $|V_{us}|$ using our preliminary result of $f_+(0)$ with 
the previous results~\cite{Boyle:2013gsa,Boyle:2015hfa,Aoki:2017spo,Carrasco:2016kpy,Bazavov:2018kjg,Ishikawa:2022otj}.
$|V_{us}|$ determined from the $K_{\ell 2}$ decay are also
plotted using $F_K/F_\pi$ of our preliminary result and 
PDG22~\cite{ParticleDataGroup:2022pth}.
The closed (open) symbols represent results 
in the continuum limit (at a finite lattice spacing).
The inner and outer errors
express the lattice QCD and total errors.
The total error is evaluated by adding the errors in the lattice QCD and experiment in quadrature.
The value of $|V_{us}|$ determined from the unitarity of the CKM matrix using $|V_{ud}|$ in Refs.~\cite{Seng:2018yzq} and \cite{Hardy:2020qwl} are presented by the light blue and gray bands, respectively.
  \label{fig:f0_vus}
 }
\end{figure}

In the left panel of Fig.~\ref{fig:f0_vus}, our preliminary result of 
$f_+(0)$ is compared with the previous calculations in 
Refs.~\cite{Dawson:2006qc,Boyle:2007qe,Lubicz:2009ht,Bazavov:2012cd,Boyle:2013gsa,Bazavov:2013maa,Boyle:2015hfa,Carrasco:2016kpy,Aoki:2017spo,Bazavov:2018kjg} including our previous one~\cite{Ishikawa:2022otj}.
Our result is reasonably consistent with the previous results within 
about 2 $\sigma$, although systematic errors in our result are 
not estimated yet.

Combining our $f_+(0)$ and
the experimental value, 
$|V_{us}|f_+(0) = 0.21654(41)$~\cite{Moulson:2017ive},
the value of $|V_{us}|$ is estimated.
Our result is plotted in the right panel of Fig.~\ref{fig:f0_vus}
together with the previous results~\cite{Boyle:2013gsa,Boyle:2015hfa,Aoki:2017spo,Carrasco:2016kpy,Bazavov:2018kjg,Ishikawa:2022otj}.
Again, our result reasonably agrees with the previous ones
determined from $f_+(0)$ through the $K_{\ell 3}$ decay process.
The figure also shows that our preliminary result is consistent
with $|V_{us}|$ obtained from the $K_{\ell 2}$ decay
using $|V_{us}|F_K/|V_{ud}|F_\pi = 0.27599(37)$~\cite{Moulson:2017ive}
and the ratio of the decay constants $F_K/F_\pi$.
In the figure, we plot $|V_{us}|$ using $F_K/F_\pi$
in PDG22~\cite{ParticleDataGroup:2022pth} and also our preliminary result
of $F_K/F_\pi$ in the continuum limit calculated with the PACS10 configurations.
In contrast to the consistencies, our preliminary result differs
from the estimated value of $|V_{us}|$ through the unitarity of the CKM matrix
using $|V_{ud}|$ by 2--3 $\sigma$ depending on 
the error of $|V_{ud}|$~\cite{Seng:2018yzq,Hardy:2020qwl}.
To clarify the discrepancy, it is important to decrease the uncertainties
in the lattice QCD calculations and also the ones in the $|V_{ud}|$ 
determinations.

\section{Summary}

We have presented the preliminary result of the $K_{\ell 3}$ form factors
calculated with the PACS10 configuration
at the lattice spacing of $a = 0.041$ fm, where the pion and kaon
masses are physical ones and the spatial extent is more than 10 fm.
The form factors are calculated using the local and conserved vector
currents.
We have observed they have different lattice spacing dependence.
Using the results calculated on the other two PACS10 configurations at
the coarser lattice spacings,
we have obtained the preliminary result of $f_+(0)$ in the continuum limit.
The value of $|V_{us}|$ estimated using our preliminary $f_+(0)$
reasonably agrees with the previous results determined through 
the $K_{\ell 3}$ decay and also the ones through the $K_{\ell 2}$ decay.
On the other hand, our preliminary result differs from the ones
through the CKM unitarity using $|V_{ud}|$ by 2--3 $\sigma$.
In future to finalize our calculation, we will estimate systematic errors 
of $f_+(0)$, {\it e.g.,} fit forms for a $q^2$ interpolations and continuum extrapolations, isospin breaking effects, and a dynamical charm quark effect. 
Furthermore, we plan to evaluate the phase space integral of the $K_{\ell 3}$ 
decay from our form factors as in our previous work.

\section*{Acknowledgments}
Numerical calculations in this work were performed on Oakforest-PACS
in Joint Center for Advanced High Performance Computing (JCAHPC)
under Multidisciplinary Cooperative Research Program of Center for Computational Sciences, University of Tsukuba.
This research also used computational resources of Oakforest-PACS
by Information Technology Center of the University of Tokyo,
and of Fugaku by RIKEN CCS
through the HPCI System Research Project (Project ID: hp170022, hp180051, hp180072, hp180126, hp190025, hp190081, hp200062, hp200167, hp210112, hp220079, hp230199).
The calculation employed OpenQCD system\footnote{http://luscher.web.cern.ch/luscher/openQCD/}.
This work was supported in part by Grants-in-Aid 
for Scientific Research from the Ministry of Education, Culture, Sports, 
Science and Technology (Nos. 19H01892, 23H01195) and
MEXT as ``Program for Promoting Researches on the Supercomputer Fugaku'' (Search for physics beyond the standard model using large-scale lattice QCD simulation and development of AI technology toward next-generation lattice QCD; Grant Number JPMXP1020230409).
This work was supported by the JLDG constructed over the SINET5 of NII.

\bibliography{reference}

\end{document}